\documentclass[twocolumn,floatfix,pre,showpacs]{revtex4-1}
\pdfoutput=1
\usepackage{graphicx}
\usepackage{amssymb}
\usepackage{amsmath}
\usepackage{amsthm}
\usepackage{colonequals}
\usepackage{soul}
\usepackage{cancel}
\usepackage{color}

\begin{document}
\title{Collective dynamics of dipolar and multipolar colloids: from passive to active systems}
\author{Sabine H.~L.~Klapp}
\email{klapp@physik.tu-berlin.de}
\affiliation{
Institut f\"ur Theoretische Physik, Sekr.~EW 7--1,
Technische Universit\"at Berlin, Hardenbergstrasse 36,
D-10623 Berlin, Germany
}
\date{\today}

\begin{abstract}
This article reviews recent research on the collective dynamical behavior of colloids with dipolar or multipolar interactions. Indeed, whereas 
equilibrium structures and static self-assembly of such systems 
 are now rather well understood, the past years have seen an explosion of interest in understanding dynamicals aspects, from the relaxation dynamics
 of strongly correlated dipolar networks over systems driven by time-dependent, electric or magnetic fields, to pattern formation and dynamical
 control of active, self-propelled systems. Unraveling the underlying mechanisms is crucial for a deeper understanding of 
self-assembly in and out of equilibrium and the use of such particles as functional devices.
At the same time, the complex dynamics of dipolar colloids poses challenging physical questions and puts forward their role as model systems
for nonlinear behavior in condensed matter physics. Here we attempt to give an overview of these developments, with an emphasis on theoretical and simulation studies.
\end{abstract}
\maketitle

\section{Introduction and scope of this article}
Colloids with anisotropic, directional interactions play nowadays a major role in the field of self-assembly of colloidal matter, in microfluidics, the design of functional
devices such as robots and sensors, but also as theoretically and experimentally accessible model systems in condensed matter physics.  
A paradigm example are dipolar colloids whose interactions are governed by permanent or field-induced, magnetic or electric dipole moments, as well as particles 
with more complex multipolar interactions. 

While earlier research has rather focused on understanding the (often unusual) equilibrium phase behavior and static self-assembly of such systems, 
the last years have seen substantial progress in understanding dynamical properties \cite{Dobnikar2013}, 
from the single-particle response over the collective dynamics
of strongly correlated, driven systems \cite{Yan2012,Klapp_News} towards the dynamics of active dipolar systems \cite{Snezhko2011,Palacci2013}.
The purpose of the present review is to give an overview of recent developments in this emerging field from a theoretical point of view,
and to outline perspectives for future research. The focus lies on systems composed of {\em spherical} particles with dipolar or multipolar interactions since these have been studied in most detail so far.
However, a trend towards shape-anisotropic dipolar systems is already foreseeable
\cite{Bharti2015,Tierno_review}.

We start by discussing dynamical properties close to equilibrium, such as 
the relaxation dynamics and gelation of self-assembled structures. These properties are highly relevant, e.g., for the resulting materials' elastic response and conductivity \cite{Ilg2011}.
A second topic is the {\em non-equilibrium} behavior of dipolar systems generated by a time-dependent, rotating external field. Indeed,
time-dependent fields have recently shown to induce not only unusual (quasi-static) structures, but also complex nonlinear dynamics such as synchronization (and related structural)
transitions \cite{Yan2012}. Third, we discuss the collective (translational) transport properties of dipolar systems in alternating fields \cite{Juniper2015} and in "active" systems 
where the particles are driven by an an internal energy source \cite{Palacci2013}. Indeed, active dipoles are an exemplary topic where current
research on active-particle systems and that on passive complex systems meet, and where a stimulating interplay can be foreseen. In fact, there are many research
themes which are "hot topics" in both areas, such as the interplay of clustering/aggregation and equilibrium phase separation \cite{Schwarzlinek2012,Redner2013}, as well
as the control of (single-particle and collective) motion by external, magnetic or electric fields \cite{Snezhko2011}. Thus, a comprehensive discussion highlighting the interface of these fields of research is timely.

There are a number of topics which are related to the overall theme of this article but are not covered or touched only briefly here. Examples are 
the equilibrium structures of dipolar colloids in the ground state and at finite temperatures (see, e.g. \cite{Holm2005}), the 
behavior of electro- and magnetorheological systems in static fields \cite{Vicente2011}, the dynamics under shear flow \cite{Ilg_Odenbach}, the behavior of magnetic elastomers \cite{Menzel_review,Ilg_review}, and the dynamics and growth
of the closely related patchy particle systems \cite{Sciortino2011,daGama2014}.
\section{Models}\label{sec:model}
\begin{figure}%
	\centering
	\includegraphics[width=\linewidth]{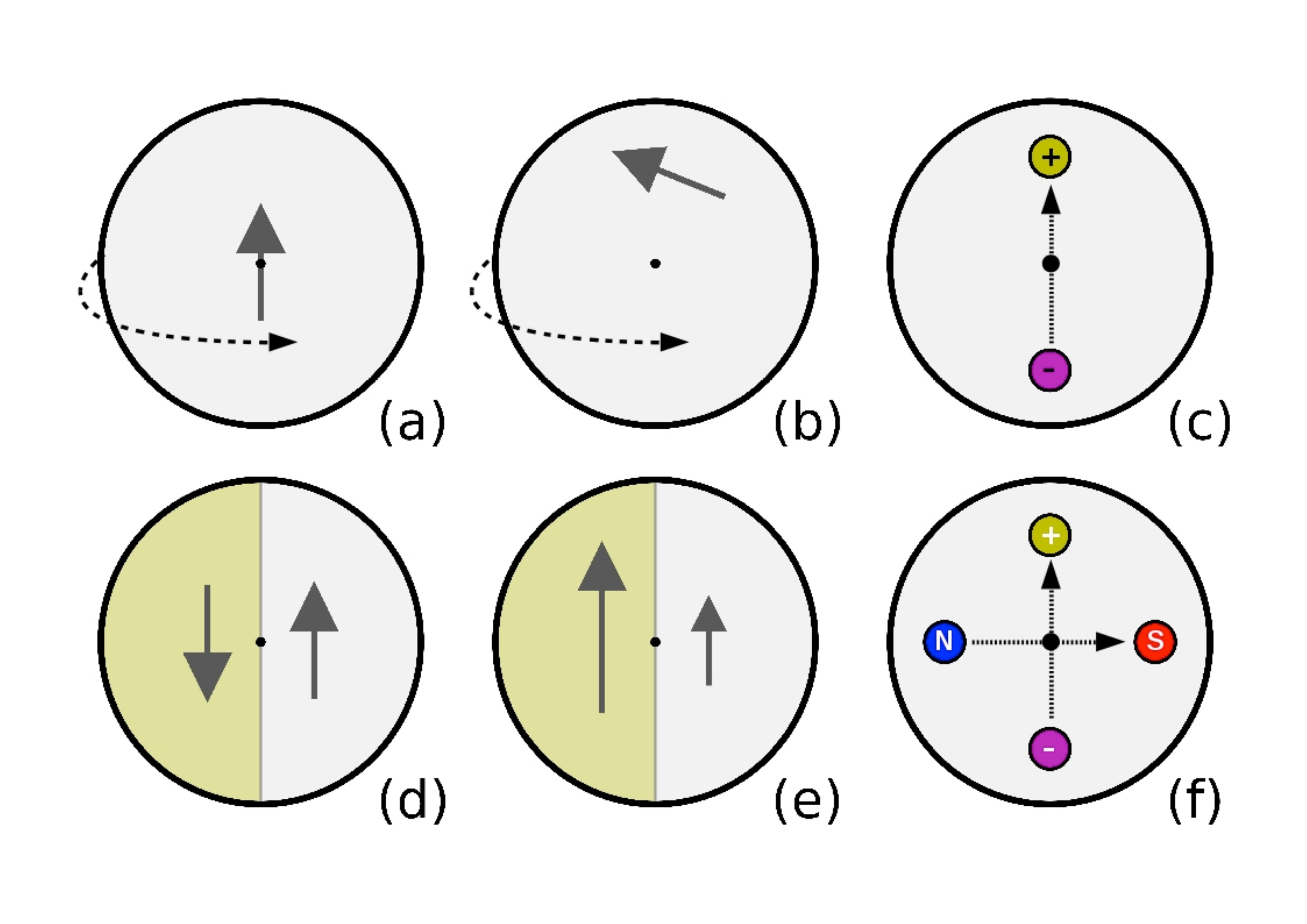}
	\caption{Sketches of model colloids with dipolar (a)-(c) or multipolar (d)-(e) character. Specifically: (a) Dipolar hard or soft sphere with permanent point dipole moment in its center, (b) sphere with off-centered point dipole shifted laterally (dipole moment is perpendicular to the radius vector), (c) sphere with two charges representing an extended (permanent or induced) dipole.
(d) Janus-like sphere with two induced, off-centered point dipoles oriented in opposite direction, (e) Janus-like sphere with two induced, parallel dipoles, (d) sphere with two crossed, electric and magnetic, dipoles
induced by bidirectional fields.}	
	\label{models}
\end{figure}

This section gives an overview of models typically discussed in the context of dipolar colloidal systems.
We focus here on monodisperse systems of {\em spherical} particles. The 
directional interaction between particles $i$ and $j$, $U^{\mathrm{aniso}}(ij)$, results either from permanent magnetic (ferromagnetic) moments or charge
distributions, respectively, or it is induced by an external field. In addition, the spheres interact via a short-range potential $U^{\mathrm{sr}}(ij)$, such 
as the purely respulsive hard-sphere and soft-sphere potentials, or the Lennard-Jones potential which includes attractive (van-der-Waals) interactions.
Some important representatives of such particles are shown in Fig.~\ref{models}.
The "paradigm" model for ferromagnetic particles (permanent moment) are hard spheres with embedded point dipoles $\boldsymbol{\mu}_i$ and
$\boldsymbol{\mu}_j$ (the so-called "dipolar hard sphere" (DHS) model) in their center, Fig.~\ref{models}(a),
where the anisotropic part of the interaction is given by the usual dipole-dipole potential,
\begin{equation}
    \label{eq:interaction}
    U^{\mathrm{D}}({\bf r}_{ij}, \boldsymbol{\mu}_i, \boldsymbol{\mu}_j) = 
\frac{\boldsymbol{\mu}_i \cdot 
    \boldsymbol{\mu}_j}{r_{ij}^3}     
    - \frac{3 ({\bf r}_{ij} \cdot \boldsymbol{\mu}_i) ({\bf r}_{ij} \cdot 
    \boldsymbol{\mu}_j)}{r_{ij}^5},
\end{equation}
with ${\bf r}_{ij}$ being the connecting vector and $r_{ij}=|{\bf r}_{ij}|$. Figure \ref{models}(b) shows a variant characterized by an {\em off-centered}, laterally shifted 
permanent point dipole; this (and related off-centered) model(s) have been recently introduced 
\cite{Abrikosov2013,Novak2015,Kantorovich2011,Yener2015}
to describe the behavior of chemically heterogeneous spherical particles such
as magnetic Janus spheres composed of two different magnetic materials (see, e.g., \cite{Sacanna2012,Yan2015_b}). 
In some cases the point dipole approximation has found to be inappropriate or computationally inefficient, thus, models with spatially separated
charges (see Fig.~\ref{models}(c)) interacting either with true Coulombic interactions \cite{Blaak2007,Miller2009,Ilg2011} or exponentially screened, Yukawa-like potentials
\cite{Schmidle2012} are used as well.

The bottom row of Fig.~\ref{models} shows models of particles with "multipolar" interactions, involving either more than one (point) dipole moment, or more than two spatially separated
charges. The models in Fig.~\ref{models}(d) and (e) have been inspired by metallodielectric Janus spheres
such as polystyrene colloids with gold patches. Experimentally, such particles have been extensively investigated \cite{Gangwal2008,Gangwal2008_b,Gangwal2010} in a quasi-2D set-up 
(realized by two confining glass plates), where they are dissolved in water and
exposed to an in-plane AC electric field. The latter creates an induced dipole moment in both parts of the metallodielectric particles.
Interestingly, not only the magnitude of these dipoles but also their direction (and thus, the character of the resulting interaction) depends 
on the frequency ($f$) of the AC field, the physical reason being the frequency dependence of the polarizability 
of the two particle domains. In particular, while the gold patch is strongly polarized along the field at any $f$,
the polarizability of the dielectric part (together with its
counterionic atmosphere) switches its sign from positive, Fig.~\ref{models}(e), to negative, Fig.~\ref{models}(d),
at a critical frequency $f_c$. Corresponding models have been suggested in \cite{Schmidle2013,Kogler2015_a}.
The resulting anisotropic interaction between two spheres is then the sum of the dipole-dipole interactions, Eq.~(\ref{eq:interaction}),
between each moment (polarization effects have so far not been investigated).
Finally, Fig.~\ref{models}(f) sketches a particle with crossed, extended dipoles induced by a combination of two external fields
\cite{Bharti2015,Kogler2015_b}. Even more complex charge distributions have been suggested in the context of "inverse patchy colloids" \cite{Bianchi2011,Bianchi2014}.
\subsection*{Common theoretical and computational methods}
Many of the theoretical studies on passive dipolar systems involve particle-resolved computer simulations such as Molecular Dynamics (MD), that is, solution of the coupled Newtonian equations
of motion; Langevin Dynamics (LD), which is based on underdamped stochastic differential equations with friction and white noise, and Brownian Dynamics (BD) based on
overdamped Langevin equations. Combinations of MD and the Navier-Stokes equation to incorporate hydrodynamic flows have been used as well
\cite{Piet2013}. Besides particle-resolved computations,
the dynamics of dipole-coupled colloids has been investigated by a variety of effective single-particle theories (to which we will refer in the text) and field methods such as (dynamical) density functional theory. The latter is based on a generalized diffusion equation, where
particle interactions are incorporated adiabatically via a free energy density functional derived from equilibrium (static) density functional theory (see, e.g., \cite{Archer2004}).
Similar to the passice case, current theoretical studies of active particle systems heavily involve particle-based simulations such as BD, but also methods incorporating hydrodynamics at low
 Reynolds numbers where the Navior-Stokes equation reduces to the Stokes equation. Common representatives are multi-particle collision dynamics (see, e.g., \cite{Goetze2010,Zoettl2014}) and Stokesian Dynamics \cite{Evans2011}. At the same
 time, these systems are studied on the basis of kinetic (Fokker-Planck-type) equations \cite{Saintillan_review}, 
 by dynamical density functional theory \cite{Menzel2016} as well as via continuum approaches \cite{Marchetti_review}. 
\section{Aggregation and relaxation dynamics}
\label{aggregation_relaxation}
In this section we summarize recent theoretical work on the dynamics of dipolar and multipolar colloids close to thermodynamic equilbrium, that is, in the absence of a driving field
or an intrinsic propulsion mechanism. In particular, we discuss aspects of kinetic aggregation and dynamical slowing-down.

\subsection{Simple dipolar systems: self-diffusion and gelation}
\label{diffusion}
Dipolar particles are prototypes of self-assembling systems: even for the simplest systems, that is, spheres with centered point dipoles (see Fig.~\ref{models}(a)),
the resulting interaction given in Eq.~(\ref{eq:interaction}) is characterized by strong anisotropy favoring head-to-tail ordering. 
Equilibrium aspects of the resulting cluster- and chain formation in model systems with dipolar (or multipolar) interactions have been studied
extensively by computer simulations, ground state calculations, and association theories
(see e.g., \cite{Holm2005,Rovigatti2011,Kantorovich2015} and references therein)
and are
nowadays quite well understood. This concerns both, systems in external fields and zero-field systems involving permanent dipoles.
We note, however, that establishing the precise relation between clustering and the equilibrium {\em phase diagram}, particularly
the existence of a first-order vapor-liquid transition predicted by perturbation theories, 
has been a challenge for decades even for the paradigmatic DHS model \cite{Rovigatti2011}.

The strong tendency of dipolar spheres to self-assemble into clusters and chains prompts the question whether simple dipolar particles can form {\em gels}, similar to what is
found in systems of "patchy" particles \cite{Sciortino2011}. Patchy particles interact via short-range attractive potentials between a finite number of interaction sites on each particle. This leads
to long-lived physical bonds forming eventually a system-spanning network. The latter is a prerequiste for a dynamically arrested state with broken ergodicity.
In conventional, DHS-like, models
particles arrange in clusters and chains which, however, appear not sufficiently connected to form gels. Therefore, a number of recent MD simulation studies \cite{Blaak2007,Miller2009,Ilg2011} 
have considered dumbbell-like particles made up of two interpretating soft spheres carrying oppositie charges $\pm q$ at their centres,
separated by a fixed distance $d$. The resulting dipole moment is $\mu=qd$ (see Fig.~\ref{models}(c) for a sketch of corresponding spherical particle).
The extended dipole enhances string formation and
facilitates branching of the chains, leading to true threedimensional (3D) networks. Dynamical signatures of this process are, e.g., the emergence of sub-diffusive ranges
in the mean-squared displacement (MSD) 
$\Delta {\bf r}^2(t)=\left\langle N^{-1}\sum_{i=1}^N\left({\bf r}_j(t)-{\bf r}_j(0)\right)^2\right\rangle$, where $t$ is the time. Sub-diffusion is characterized 
by a time-dependence of the MSD as $t^\alpha$ with $\alpha<1$. At the same time, one observes
a marked slowing down of the intermediate scattering function 
$F_s({\bf q},t) =\left\langle N^{-1}\sum_{j=1}^N\exp\left[i{\bf q}\cdot{\bf r}_j(t)-{\bf r}_j(0)\right]\right\rangle$,
yielding a finite non-ergodicity parameter
$F_s({\bf q},t\to\infty)$
\cite{Blaak2007,Miller2009,Ilg2011}. Overall, the behavior
is thus similar to that observed in short-ranged attractive systems forming reversible gels \cite{Sciortino2011}. Moreover, the dumbbell systems shows interesting nonlinear response to an external
magnetic field \cite{Ilg2011}.

We further note that some aspects of gel-like dynamics, particularly sub-diffusion and dynamically heterogeneous behavior, has also been observed in a MD simulation study
of strongly coupled dipolar soft spheres in a homogeneous external field \cite{Jordanov2011}. 
The latter "helps" the particles in forming extended structures, that is, chains along the field, which finally percolate
in the field direction. Interestingly, these systems additionally display evidence of {\em super-diffusion} (characterized by a MSD $\propto t^{\alpha}$ with $\alpha>1$) at larger times,
consistent with dynamic light scattering experiments  \cite{Mertelj2009} of such strongly coupled systems. 
On the other hand, slow dynamical relaxation related to static percolation has also been predicted to occur
in a Langevin dynamics study of a cobalt nanoparticle suspension \cite{Sreekumari2013}, as well as in Discontinuos Molecular Dynamics simulations
of a 2D system of spheres with short-ranged dipole-like
interactions \cite{Schmidle2012}.
\subsection{Multipolar particles: Networks and aggregation}
\label{networks}
Advances in material chemistry have enabled fabrication of novel particle systems
where the anisotropic interparticle interactions can not be approximated by those between dipoles, but have rather multipolar character. 
In many cases, these interactions are induced by external electric or magnetic fields (for recent reviews
from the experimental side, see \cite{Bharti2015,Kretzschmar2011}).
\subsubsection*{Multipolar particles in uni-directional fields}
 A prime example are patchy metallodielectric particles such as polystyrene particles with
gold patches. In a unidirectional field, these particles (for sketches see Fig.~\ref{models}(d) and (e)) can form a variety of novel structures, such
as zig-zag chains along the field (also reported for magnetic Janus spheres \cite{Yan2015_b}) and, importantly, strings and clusters {\em perpendicular} to the field \cite{Gangwal2010}.
 
\begin{figure}%
	\centering
	\includegraphics[width=\linewidth]{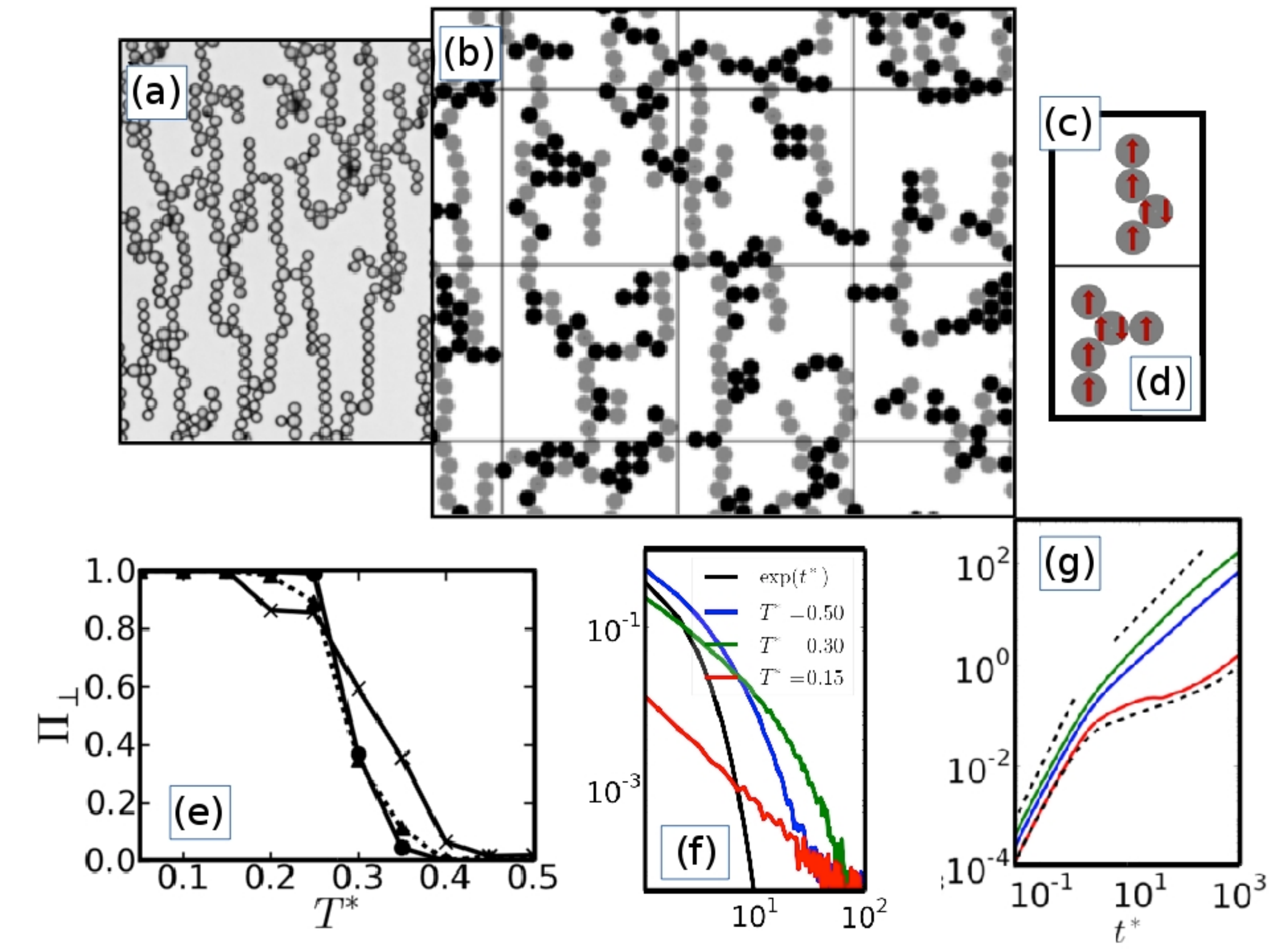}
	\caption{Formation of crosslinked 2D networks in systems of metallodielectric particles (mixed with simple dielectric particles) in an uniaxial electric field \cite{Schmidle2013}.
	(a) Optical microscopy images of the experimental system of 5 $\mu$m particles, (b) MC simulation snapshot of a corresponding model (black: multipolar particle, see Fig.~\ref{models}(d), 
	grey: simple dipolar particle, Fig.~\ref{models}(a))
	at a temperature below the transversal percolation threshold, (c) and (d) ground state structures of small clusters illustrating energetically preferred configurations, (e)
	transversal percolation probability as function of reduced temperature $T^{*}$ for different system sizes, revealing a transversal percolation transition. 
	The longitudinal percolation transition occurs at somewhat
	larger $T^{*}$. (f) Bond correlation function and (g) mean-squared displacement (transversal direction) as functions of time ($t^{*}$). In (g), the lowest curve pertains to a temperature
	below transversal percolation.}
	\label{aggregation}
\end{figure}

The latter phenomenon, which occurs at high frequencies $f>f_c$ of the applied AC field, was studied theoretically in \cite{Schmidle2013} by means of 
MC and MD simulations. The model consists of a binary mixture of spheres with two laterally shifted, oppositely oriented, dipole moments (see Fig.~\ref{models}(d)),
and spheres with only one dipole moment (representing "defect" particles in real systems). All dipole vectors are orientationally fixed due to their induced character. The simulations demonstrate formation of 
percolated, gel-like structures both parallel (longitudinal) and perpendicular (transversal) to the field,
consistent with optical microscopy experiments on metallodielectric
particles \cite{Gangwal2010} (see Fig.~\ref{aggregation}(a)-(b)). The transformation between these structures takes
place via two clear, longitudinal and transverse percolation
transitions, with the novel 2D-crosslinked structure being stable for a broad
range of concentrations and coupling strengths \cite{Schmidle2013}. Moreover, the networks are
 characterized by strongly hindered translational dynamics: The MSD shows a plateau in transversal direction after the second percolation
 transition. Further, orientational correlations reveal an extremly long bond lifetime, suggesting a strong persistence of the network (see Fig.~\ref{aggregation}(f)-(g)).
 Similar dynamic behavior has been very recently reported in MD many-particle simulations of "inverse patchy colloids" characterized by heterogeneously charged
 surfaces \cite{Ferrari2015}: These systems do not only self-assemble in a variety of structures depending on the charge distribution \cite{Bianchi2014}; they also display a drastic
 decrease of the diffusion coefficient with decreasing temperature or increasing density, indicating again the existence of a gel-like state \cite{Ferrari2015}. 
From an applicational point of view, the gelation is of prime importance for the system's elastic response and its transport properties.

Completely different behavior of metallodielectric spheres occurs at certain frequencies below the critical one: Here, the 
particles do not only interact in a different manner; they also display {\em self-propulsion}
in directions perpendicular to the field \cite{Gangwal2008_b}. The collective dynamics of the resulting "dipolar swimmers" is discussed in Sec.~\ref{transport}.
\subsubsection*{Multipolar particles in bi-directional fields}
Another recent topic is the aggregation behavior of polarizable colloidal spheres under the influence of {\em bi-directional} fields. An exemplary system
are latex spheres with embedded paramagnetic nanoparticles; these spheres are responsive to both, electric and magnetic fields. Using fields in an orthogonal set-up the particles
have been shown to build 2D, percolated networks, whose structure appears to depend on time and 
whose topological (connectivity) properties 
can be tuned independently by the two fields \cite{Bharti2015}. In \cite{Kogler2015_b},
Brownian Dynamics simulations were carried out on the basis of a simplified, yet generic model suitable for (spherical) particles in crossed fields. The model contains four charges with variable separation $\delta$ (see Fig.~\ref{models}(f)), which interact via screened (thus, short-ranged) Yukawa potentials. At very low temperatures the model predicts non-equilibrium aggregation into large clusters
with square- or hexagonal-like local order. The overall behavior (including the cluster's fractal dimension and behavior of bond time correlations) is similar to diffusion-limited aggregation.
Upon increasing the temperature the clusters "melt" and the system transforms into a fluid phase, consistent with results of a simple mean-field density functional theory \cite{Kogler2015_b}.

Importantly, dynamical clustering is a "hot topic" not only in the area of {\em passive} colloids discussed so far,
but also in systems of {\em active}, self-propelled colloids (see also Sec.~\ref{active}). Experiments \cite{Theurkauff2012} and 
simulations \cite{Mognetti2013,Pohl2014}
involving (chemically driven) colloidal swimmers
report formation of a novel cluster phase (resulting from a permanent dynamical particle merging and separation) at intermediate densities. 
Contrary to passive system, clusters can already occur in {\em purely repulsive} active systems due to a self-trapping mechanism \cite{Speck2014}.
An intriguing question for both, 
passive \cite{Kogler2015_b} and active colloids \cite{Schwarzlinek2012,Redner2013}, concerns the relation between clustering and equilibrium phase separation. Here we may expect
that the two fields can strongly benefit from one another, especially when active colloids with directional interactions are considered \cite{Kogler2015_a,Kaiser2015}.
\section{Collective behavior in rotating fields}
\label{Sec_rotating}
We now turn to systems driven {\em out of equilibrium} 
by means of rotating, magnetic or
electric, external fields (see \cite{Martin_review} and \cite{Tierno_review}, respectively, for 
recent reviews covering the experimental and material science perspective).

Earlier experimental and theoretical studies in the area of rotating fields typically focus on the field-induced
dynamics of an {\em isolated} colloid such as a magnetic
rod \cite{McNaughton2006,Tierno2009,Coq2010,Dhar2007,Cimurs2013}, a magnetic chain \cite{casic} or
filament \cite{Dreyfus2005}, or an optically excitable nanorod \cite{shelton} in
a viscous medium. Understanding the resulting single-particle rotational
dynamics is important for the advancement of actuators \cite{Coq2010}, sensors \cite{McNaughton2006}, molecular
switches, particles in optical traps \cite{shelton}, and in the more general
context of microfluidics \cite{Dhar2007}.
%

However, even more intriguing is the {\em collective}, dynamical self-assembly
of colloids exposed to rotating fields. Indeed, in
material science, rotating fields have been realized
as a powerful tool to control self-assembly processes into functional materials
 \cite{Yan2015_b,leunissen}. In fact, even relatively simple dipolar systems can form novel structures, which do not exist in equilibrium.

\subsection{Polarizable spheres in rotating fields}
A classical example in this context,
first discussed by Martin {\em et al.} \cite{martin2} are 3D systems of
paramagnetic (or polarizable) spherical particles in magnetic (electric) fields
rotating in a plane (biaxial field). Specifically, let us assume a field ${\bf B}(t)$ rotating with frequency
$\omega_0$ in the $x$-$y$-plane,
\begin{equation}
{\bf B}(t) = B_0 ( {\bf e}_x \cos \omega_0 t + {\bf e}_y \sin \omega_0 t ).
\end{equation}
Since the particles are polarizable and spherical, they aquire an induced dipole moment parallel to the field
at {\em all} times, that is
\begin{equation}
\label{induced}
    {\boldsymbol{\mu}}_i (t) = {\boldsymbol{\mu}}_j (t)
    = \mu ( {\bf e}_x \cos \omega_0 t + {\bf e}_y \sin \omega_0 t )
\end{equation}
with $\mu\propto B_0$. Note that this simple dipole-field relation does not hold for the induced dipole moment of non-spherical paramagnetic
particles characterized by different polarizabilities along and perpendicular to the symmetry axes; here, the direction of
${\boldsymbol{\mu}}_i (t)$ can deviate from that of the field (see, e.g. \cite{Cimurs2013}).

For sufficiently large field strength $B_0$, both experiments and
computer simulations \cite{martin1,martin2,martin6,elsner,smallenburg2010} reveal the formation of
layers in the field plane, \textit{i.e.}~a {\em spatial} symmetry breaking
induced by the rotating field. The physical mechanism becomes clear by considering the
time-averaged dipolar potential \cite{martin1} defined as
\begin{multline}
    \label{eq:dipole_avg}
    U^{\mathrm{ID}} ({\bf r}_{ij}) = \tau^{-1} \int_{t_0}^{t_0 + \tau}
    U^\mathrm{D}({\bf r}_{ij}, \boldsymbol{\mu}_i(t), \boldsymbol{\mu}_j(t))
    d t \\
    = - \mu^2 \frac{(1 - 3 \cos^2 \Theta_{ij})}{2 r^3_{ij}} .
\end{multline}
In this equation, $U^\mathrm{D}$ is the dipole-dipole potential 
Eq.~(\ref{eq:interaction}) between the induced dipoles given in Eq.~(\ref{induced}),
$\tau = 2 \pi / \omega_0$ is the oscillation period,
and $\Theta_{ij}$ is the angle between the interparticle vector ${\bf r}_{ij}$
and the direction perpendicular to the plane of the field (i.e., the $z$-direction). 
The time-averaged potential corresponds to
an inverted dipolar (ID) potential, which is attractive if the angle
$\Theta_{ij}$ satisfies $\cos^2 \Theta_{ij} < 1/3$, \textit{i.e.}~if the
particles $i$ and $j$ are approximately in the same plane with respect to the
field. Conversely, if the angle $\Theta_{ij}$ satisfies $1/3 < \cos^2
\Theta_{ij}$, the particles repel each other. This combination of in-plane
attraction and repulsion along the rotation axis \cite{martin1,elsner}
explains why layers are a favorable configuration.

From a theoretical perspective, the above time-averaged potential allows for an "effective equilibrium description" of the rotating-field-driven dipolar fluid.
It should be noted, however, that the resulting overall (quasi-equilibrium) phase behavior additionaly depends 
on the remaining parts of the effective pair interaction. Using MC simulations, Smallenburg and Dijkstra \cite{smallenburg2010} have 
shown that layered-fluid phase do appear for charged spheres with inverted dipolar interactions, Eq.~(\ref{eq:dipole_avg}). On the other hand,
the phase diagram of colloidal hard spheres with inverted dipolar interactions displays a gas-liquid transition, a hexagonal ABC stacked crystal phase, a stretched hexagonal-close-packed crystal,
but no layered structures.

Another intriguing example of the use of rotating fields to create a specific time-averaged (quasi-equilibrium) interaction
has been proposed by Osterman {\em et al.} \cite{osterman2009} who considered superparamagnetic spheres in a
precessing magnetic field. Here, a static and a rotating field are combined in a "magic" opening angle ($\theta=\arctan(1/\sqrt{3}\approx 54.7^0$), such that
the spheres effectively feel an isotropic pair attraction similar to the van der Waals force between atoms. However, additional
polarization interactions, and thus, many-body interactions, lead to the formation of crosslinked network, which coarsen and eventually form  
"self-healing" membranes. These investigations have been recently extended towards 2D systems \cite{Mueller2014}: depending on the opening angle $\theta$, a broad range
of quasi-equilibrium structures from hexagonal crystals to froth-like patterns is observed both in experiments, and in parallel numerical (MC) simulations of a model system involving a mixture of patchy and non-patchy particles. This study \cite{Mueller2014} underlines the role of many-body effects for structure formation in complex paramagnetic systems.
\subsection{Permanent dipoles in rotating fields: Layering and synchronization}
As discussed above, many phenomena in rotating-field driven systems of induced dipoles can be explained from an
equilibrium perspective involving the free energy and resulting phase behavior
in the time-averaged field. 
Less is
known about the corresponding behavior of particles with \textit{permanent}
dipole moments, such as the (ferromagnetic) particles of a ferrofluid. Here,
the individual orientations can be different from the one of the rotating field due to thermal effects
and effects from the solvent. Thus, the essential prerequisite for building a time-averaged potential, that is, full synchronization with the field (see Eq.~(\ref{induced})),
can break down. 

\begin{figure}
	\centering
	\includegraphics[width=\linewidth]{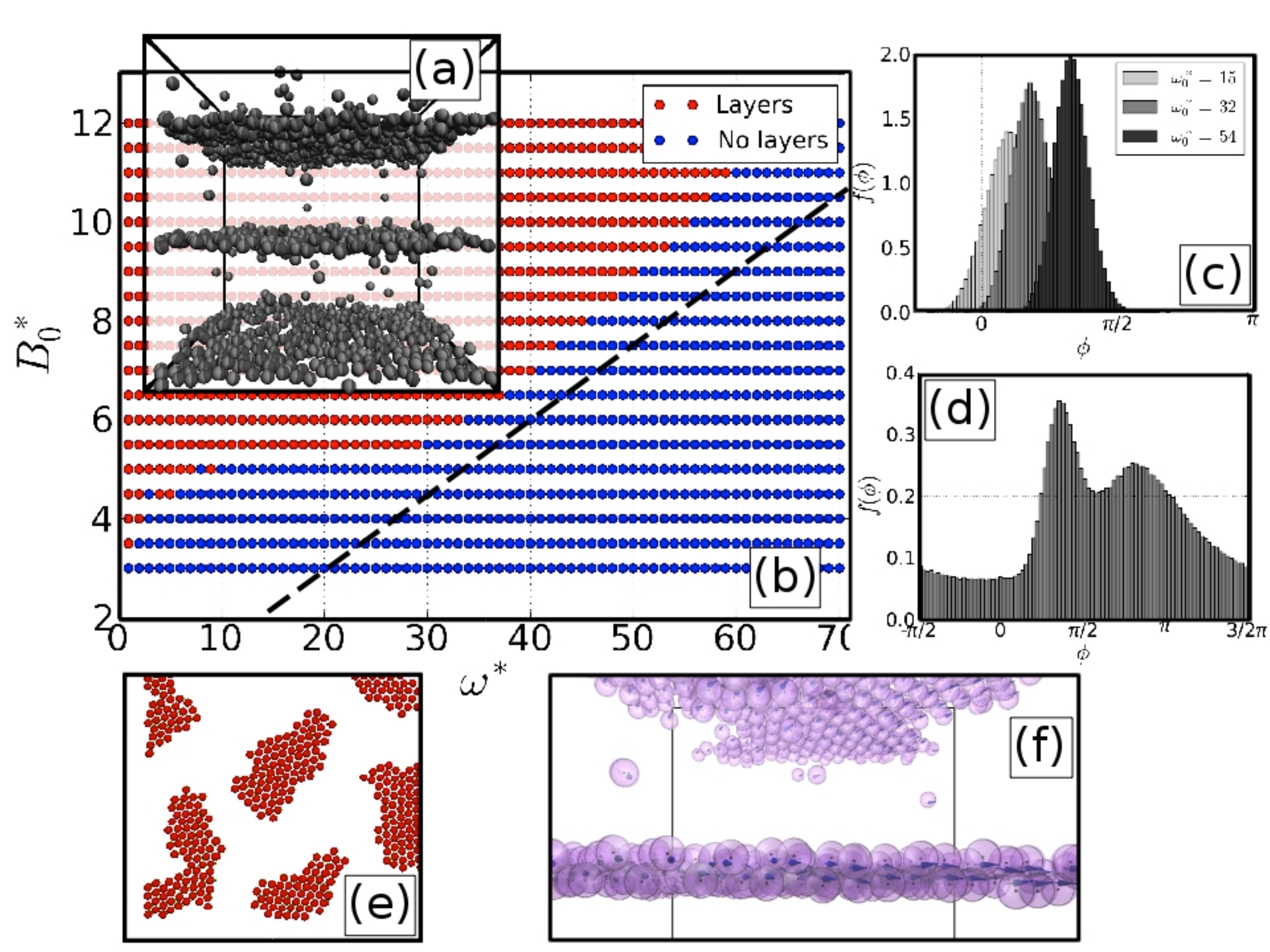}
	\caption{Structure formation in simulated systems of permanent dipoles in rotating (planar) fields \cite{Jaeger2011,Jaeger2012,Yener2015}.
	(a) Layer formation in a 3D system of centered dipoles, Fig.~\ref{models}(a). (b)
	Corresponding non-equilibrium state diagram in the field strength ($B_0^{*}$)--frequency ($\omega_0^{*}$) plane. The dashed line
	indicates the results from the effective single-particle theory, Eq.~(\ref{eq:non_linear}).
	(c) and (d): Distribution of phase differences in the layered (c) and unlayered (d) region,
	showing that layering is intimately related to synchronization of the particles \cite{Jaeger2011}. (e) Cluster formation in the synchronized regime of a 2D system of permanent (centered) dipoles in a rotating field \cite{Jaeger2012}. (f) Double layer formation in a 3D system of off-centered dipoles (see Fig.~\ref{models}(b)) \cite{Yener2015} in the synchronized regime.}
	\label{synchro}
\end{figure}

The behavior of a 3D system of permanent, strongly coupled dipoles in a planar rotating field has been investigated
in Ref.~\cite{Jaeger2011} on the basis of LD simulations and density functional theory. In this study,
a full non-equilibrium state diagram as function of the driving frequency and the field strength was mapped out, see Fig.~\ref{synchro}(b).
At small frequencies and sufficiently large $B_0$, the system is in a {\em synchronized} state, where 
the individual particles follow the field with (on average) constant
phase difference. This is indicated by a single-peaked distribution of phase differences, $f(\Phi)$, where $\Phi_i$ is the in-plane angle between dipole
vector $i$ and the external field, see Fig.~\ref{synchro}(c). Thus, the overall magnetization is 
${\bf M}(t) \approx M_0 ( {\bf e}_x \cos(\omega_0 t + \phi_0)
    + {\bf e}_y\sin(\omega_0 t + \phi_0))$ where $\phi_0$ is the average phase difference. 
The synchronization is accompanied
by 3D layer formation, similar to
systems of induced dipoles in rotating fields. At very low frequencies, the translational structure within these layers is characterized by small chains
with head-tail ordering (similar to what is seen in equilibrium dipolar systems). At somewhat larger frequencies, the local structure rather
resembles that in a isotropically interacting 2D system which tend to form hexagonal lattices.
However, for permanent dipoles the layering breaks down even at small $\omega_0$ when the field strength $B_0$
becomes too small. As shown in Ref.~\cite{Jaeger2011}
the transformation of an unlayered (small $B_0$) into a layered/synchronized (large $B_0$) state
can be described as a quasi-equilibrium transition induced by the competition between the time-averaged, inverted dipolar interaction (which favors layering)
and the loss of translational entropy related to the onset of 1D translational order. This has been demonstrated on the basis of 
an effective (equilibrium) free energy functional.

Completely different behavior is found at high frequencies and field strengths.
Under these conditions, the picture of synchronously rotating dipoles breaks down. Instead, one
observes a mixture of rotating and counter-rotating or resting particles, as
an analysis of $f(\Phi)$ reveals (see Fig.~\ref{synchro}(d)). Similar behavior has been observed in systems of rotating ellipsoids \cite{shelton} and theoretically analyzed
in Ref.~\cite{Haertel2010} by dynamical density functional theory.

The desynchronization induces, at the same time, a breakdown of the translational, layered structure.
Despite this complex many-particle behavior, the high-field boundary between layered and non-layered
states can be well described within an effective {\em single-particle} approach for
the rotational motion of a permanent dipole in a viscous medium. This dipole experiences, first,
a field-induced torque given by ${\boldsymbol{\tau}}_B={\boldsymbol{\mu}}\times {\bf B}=
\mu B_0\sin\Phi$ (where $\Phi$ is again the phase lag), and second, a viscous (dissipative) torque
${\boldsymbol{\tau}}_{diss}=\gamma d\Phi/dt$. For large friction constants $\gamma$, the motion
can be considered as overdamped, leading to a first-order (in time), nonlinear equation of motion
for the phase angle $\Phi$,
\begin{equation}
\label{eq:non_linear}
\frac{d\Phi}{d \tau} = \frac{\omega_0}{\omega_c} - \sin \Phi ,
\end{equation}
where $\omega_c = \mu B_0/\gamma$ and $\tau = \omega_c t$ (see dashed line in Fig.~\ref{synchro}(b)).
We note that this type of equation occurs in various contexts related to synchronizing systems
\cite{strogatz,pikovsky}; there it is often referred to as {\em Adler equation} \cite{Adler}.
For $0 \leq \omega_0 <\omega_c$ Eq.~(\ref{eq:non_linear}) has two fixed points ($\dot{\phi} = 0$, i.e., constant phase difference), with
the stable one given by $\phi = \arcsin (\omega_0/\omega_c)$. Here, the torque due to friction equals the torque
due to the field. At $\omega_0 = \omega_c$, \textit{i.e.}~at $\phi =
\pi/2$, the two solutions form a saddle-node bifurcation and there are no fixed
points for $\omega_0 > \omega_c$. At these high frequencies, the maximal torque
that can be exerted by the field is insufficient to balance the frictional
torque. 

The main simulation result of Ref.~\cite{Jaeger2011}, namely that the rotational dynamics of the individual particles (synchronous versus asynchronous motion) determines the structure
of a many-body system of permanent dipoles, has been recently confirmed by experiments involving magnetic Janus spheres with both, ferromagnetic and paramagnetic response \cite{Yan2015,Yan2015_b}. We come back to these studies in Sec.~\ref{rotating_Janus}.
\subsection{2D systems in rotating fields: Clustering}
Further interesting phenomena occur when the particles are translationally confined to the plane of the rotating field.
Corresponding experiments with micron-sized (paramagnetic) particles \cite{Weddemann2010} and magnetic Janus particles \cite{Yan2015_b} have revealed
the formation of cluster structures, whose internal structure is governed by hexagonal ordering and whose size varies with time.
Such clustering and coarsening has also been observed in
Langevin and MC computer simulations \cite{Jaeger2012} of a model system involving permanent dipoles (see Fig.~\ref{synchro}(e)). There, the clustering is associated
to synchronous rotational motion (similar to the layering in 3D). The simulations \cite{Jaeger2012} further suggest that the clustering 
can be related to an underlying {\em condensation} phase transition of the (quasi-equilibrium) system defined by the time-averaged dipolar potential. Indeed,
in the 2D set up considered,
the time-averaged dipolar potential (see Eq.~(\ref{eq:dipole_avg}] for the 3D case)
is purely attractive and long-ranged, 
\begin{equation}
    \label{av_2D}
    U^{\mathrm{ID}}_{\mathrm{2D}} ({\bf r}_{ij})     = - \frac{\mu^2} {2 r^3_{ij}} .
\end{equation}
yielding a first-order vapor-liquid transition in the 2D system. Inside the corresponding coexistence curve, the system forms domains of low and high density
with the domain size growing in time. Moreover, for sufficiently large coupling strengths the translational structure within the clusters is hexagonal. The thermodynamic
conditions match those for rotating fields, supporting the relation between rotating-field-induced clustering and equilibrium phase separation \cite{Jaeger2012}. 

Importantly, these results remain qualitatively unchanged when hydrodynamic interactions (HI) in a far-field approximation are taken into account \cite{Jaeger2013}.
The corresponding expressions for dipolar fluids, involving translation-translation coupling (Rotne-Prager tensor \cite{Rotne69}), translation-rotation and rotation-rotation coupling
are given in \cite{Meriguet2005}. 
As shown in Ref.~\cite{Jaeger2013}, the main effects of HI
are a) an acceleration of the cluster formation process and b) a rotation of the clusters due to hydrodynamic translation-rotation coupling. The latter effect has
also been observed in 2D experiments of magnetic Janus colloids \cite{Yan2015}. Physically, the cluster rotation can be explained by the presence of solvent-induced,
unbalanced shear forces at the edges of the cluster.
\subsection{Janus-like magnetic colloids: Self-synchronization and novel structures}
\label{rotating_Janus}
So far, our discussion of rotating-field induced structure formation in systems of permanent dipoles
was focused on "simple", spherical particles with central, permanent dipole moments. 
Novel structures
besides layers (3D) or hexagonal clusters (2D) can form when the particle's response
to the field becomes more complex. In fact, already spheres with off-centered permanent dipoles, Fig.~\ref{models}(b), yield
different structures such as colloidal double-layers \cite{Yener2015} with phase-ordered time-dependent orientations, see Fig.~\ref{synchro}(f).

An experimental system attracting much attention \cite{Klapp_News} in this context are magnetic "Janus" spheres
investigated by Yan {\em et al.} \cite{Yan2012,Yan2015,Yan2015_b}. These Janus spheres
involve two different magnetic materials, Ag and Ni, leading to both para- and ferromagnetic properties (for other types, see e.g. \cite{Tierno_review}).
When placed in a precessing magnetic field
with an opening angle $\theta$, the particles investigated by Yan {\em et al.} \cite{Yan2012} both rotate and display vertical oscillations similar to nutations of a gyroscope, with the corresponding
frequencies being tunable by $\theta$. The presence of vertical oscillations implies an {\em additional} degree of freedom ("phase variable") not present for chemically homogeneous particles. 
Thus, the particles can not only synchronize with the field, they also can synchronize {\em with each other} (phase locking), similar to networks of coupled oscillators 
(pendulums, fire flies ...) \cite{pikovsky}.
As shown in \cite{Yan2012} by a combination of experiment, theory and computer simulations, the
synchronization (which arises due to the dipolar interactions at small distances) 
leads to the formation of different types of phase-locked dipolar dimers and eventually tubular structures \cite{Yan2012}. These structures disappear when the field
conditions are changed such that the particles desynchronize. From the theory side, synchronization transitions of coupled oscillators are often described within the Kuramoto model \cite{Kuramoto1984}. As shown in Ref.~\cite{Yan2012}, it is indeed posible to describe the collective behavior
of the Janus spheres within the Kuramoto language. In particular, one can set up an effective single-particle equation resembling the Adler equation
given in Eq.~(\ref{eq:non_linear}). In addition, the authors have performed MD simulations of LJ spheres with off-cented dipoles (supplemented by self-consistent field
calculations to taken into account polarizability) supporting the experimentally observed structure formation. In summary, this study forges a link between synchronization and self-assembly, which will certainly stimulate further investigations at the interface of dynamic soft matter and nonlinear dynamics \cite{Klapp_News}. For instance, from a more abstract perspective
the suspension of magnetic Janus particles can be viewed as a network of oscillators with non-local coupling. Such networks have been predicted \cite{Abrams2004} to spontaneously form
"Chimera" states which contain spatial domains exhibiting synchronized dynamics and others that have desynchronized dynamics. Experimentally, such states have been observed only recently e.g. in systems of coupled chemical (Belousov-Zhabotinsky) oscillators \cite{Tinsley2012}. It is an intriguing question whether suspensions of complex magnetic particles
are also capable of forming Chimera states or other complex nonlinear behavior.

\section{Collective transport and pattern formation}
\label{transport}
In the preceding section we have concentrated on the collective {\em rotational} motion and associated structure formation of dipolar colloids. Another emerging area
is the directed {\em translational} motion (transport) and its controllability by external fields. 
Here we focus on two aspects: transport by ratchet-like external potentials and transport by self-propulsion. A closely related topic, not covered in the present article, is the recently discovered pattern
formation of magnetic particles at fluctuating liquid-air and liquid-liquid interfaces (for a review, see \cite{Martin_review}): 
Here the interplay of alternating magnetic fields, (ferro-)magnetic interactions and hydrodynamic interactions 
induced by local oscillations of the liquid surface yields
a number of fascinating structures such as "snakes", "asters" \cite{Snezhko2011} and swimmers \cite{Snezkho2009}. 
These structures have been extensively investigated by experiments, a phenomenological
approach (involving amplitude equations for parametric waves) and MD simulations coupled to thin-film hydrodynamics \cite{Martin_review,Piet2013}.

Similar to the other areas discussed
in this review, investigations of (collective) dipolar transport are motivated, on the one hand, by fundamental questions such as understanding Brownian motion in complex geometries, 
exploring the emerging dynamical structures in interacting systems, and, most recently,  the interplay between self-propulsion and dipolar forces. On the other hand,
the topic of dipolar transport receives increasing interest from the application side due to the
potential use of magnetic colloids e.g. in biomedical applications (as drug delivery cargo), as micropropellers in lab-on-a-chip devices, or as active microrheological probes.
\subsection{Collective transport in ratchet potentials}
\label{ratchets}
Ratchet potentials are characterized by combinations of a static, asymmetric periodic potential and an alternating field. 
In such a potential landscape, non-equilibrium fluctuations can induce net particle transport in the absence of a biasing deterministic force. 
This generic effect arises in many areas of physics and biology \cite{Reimann2002}; and it has been studied in a large variety of 
optical \cite{Grier2005}, magnetic \cite{Tierno2007,Tierno2009,Tierno2010,Gao2011} and biological systems (e.g., molecular motors, molecular sieves).

Static, magnetic periodic potentials can be created, e.g., by using ferrite garnet films \cite{Tierno2009_b} (where ferromagnetic domains with opposite magnetization direction
are aligned in stripe-like fashion) or by using a periodic arrangement of micromagnets on ``lab-on-a-chip'' devices \cite{Gao2010,Gao2011,Tahir2011}. 
An additional oscillating field is introduced by combining the static potential with a rotating
magnetic field. The latter yields a periodic increase (decrease) of the size of domains with parallel (antiparallel) magnetization, which eventually enables transport of (para-)magnetic colloidal particles.

So far, most studies in this area have been undertaken for {\em single} colloids \cite{Tierno2009_b,Tierno2010} or systems with negligible interactions. 
However, recent experiments suggest that interaction effects are important
for the transport of magnetic chains in magnetic ratchets \cite{Tierno2012} and the self-assembly of particles into patterns on magnetic lattices \cite{Yellen2005}. Only very recently \cite{Straube2014}, first theoretical steps have been undertaken to investigate ratchet-driven transport of {\em interacting} ensembles of paramagnetic colloids. In Ref.~\cite{Straube2014},
a deterministic, analytically tractable ratchet model was developed to describe the impact of attractive or repulsive, field-induced interactions in the framework of a time-averaged field. Depending on the  sign of the effective interaction one observes
the formation of stable doublets or oscillating pairs of particles, which move with constant speed. An experimental application \cite{Martinez2015} of these
cooperative transport phenomena is
the assembly and transport of straight magnetic chains and the use of magnetic ratchets 
for microsieving of non-magnetic particles. 

Another interesting effect of particle interactions has been reported for systems of paramagnetic (iron-doped polystyrene) particles
exposed to an optical potential energy landscape composed of a (symmmetric) potential, a biasing force and an oscillating field \cite{Juniper2015}. In this situation, already a single
colloid can display complex dynamics characterized by {\em dynamical mode locking}, where the average particle velocity increases step-wise
(rather than monotonically) with the external drive. This synchronization phenomenon is significantly enhanced when instead of a single colloid a flexible magnetic chain, held together by magnetic interactions, is exposed to the potential. The chain then exhibits a "breathing mode" which 
leads to caterpillar-like motion  and to stabilization of the synchronized state.
From the theoretical side, the observed effects can be nicely described in the framework of the Frenkel-Kontorova-Model \cite{Braun2010} of solid-state physics, 
a generic model for nonlinear transport and excitations in complex geometries.

We finally mention recent theoretical work on the interplay of ratchet potentials and {\em phase separation} in 2D magnetic colloidal mixtures. 
In \cite{Lichtner2015}, a binary mixture of ferromagnetic particles (modeled by spheres with classical Heisenberg spins) and non-magnetic spheres
subject to a 1D rocking ratchet potential was investigated by means of dynamical density functional theory. 
In the absence of the external potential
the system undergoes a first-order fluid-fluid
demixing transition (driven by the magnetic interactions), resulting in spinodal decomposition in certain parameter ranges \cite{Lichtner2013}. The interplay
between this intrinsic, thermodynamic instability and time-dependent external forces then leads to a novel
dynamical instability where stripes against the symmetry of the external potential form. 
Moreover, the structural transition associated to stripe formation suppresses the usual, ratchet-driven transport of particles
along the longitudinal direction \cite{Lichtner2015}.
\subsection{Pattern formation in systems of (self-)propelled, active dipolar particles}
\label{active}
The collective behavior of active, self-propelled particles or "swimmers" has received an explosion of interest during the last years (for recent reviews
from the theory perspective, see Refs.~\cite{Marchetti_review,Romanczuk_review,Cates_review,Saintillan_review}).
From the experimental side, many types
of {\em artifical} swimmers involving colloidal particles (contrary to biological swimmers such as bacteria) have been proposed, with 
different propulsion mechanism, typically based on catalytic or phoretic effects. One prominent model of such swimmers, particularly suitable for {\em spherical} 
objects such as those considered in this review, is the so-called {\em squirmer} \cite{Lighthill1952,Ishikawa2006}, that is, spheres with a pre-described tangential velocity profile on the surface which initiates the swimming. The parameters of this profile then determine whether the squirmer is a pusher or puller. Originally introduced to mimic the synchronized beating of cilia \cite{Lighthill1952,Ishikawa2006}, 
the model is now considered as generic for spherical microswimmers including, e.g., active droplets and diffusiophoretic particles \cite{Howse2007}. Particle-based simulations of squirmer suspensions at different concentrations have demonstrated that, as a result of hydrodynamic interactions between the squirmers \cite{Goetze2010}, these
system display very rich dynamical behavior including long-range polar order \cite{Evans2011} and a phase separation between a gas-like and a cluster phase \cite{Zoettl2014}.
\subsection*{Swimmers with dipolar interactions}
A significant number of artificial swimmers involves {\em magnetic} materials and thus,
exhibits magnetic dipolar properties. Recent examples of magnetic swimmers are spherical Janus particles with a magnetic cap, which can be driven
catalytically \cite{Baraban2012} or by thermophoresis \cite{Baraban2013}. Moreover, magnetic particles with anisotropic shapes such as magnetic filaments \cite{Dreyfus2005}, nanowires, spheres
with helical tails, clusters of DNA-linked paramagnetic colloids \cite{Tierno_Gol_a,Tierno_Gol_b} and magnetic "asters" \cite{Snezkho2009} can be set into motion via different magnetic fields, see \cite{Tierno_review} for a summary of recent experimental work. In fact, the external control 
or "guidance" of swimmers by magnetic fields is an important topic from an applicational point of view. Clearly, a main effect of the external field is the suppression of rotational
diffusion. 
In this context, a recent combined experimental and simulation study \cite{Palacci2013} has reported interesting {\em collective} effects of magnetically sensitive swimmers. The swimmers
(spheres with a light-activated hematite cube) interact via phoretic attraction, leading to spontaneous aggregate into "living", crystal-like clusters. The motion of these
clusters can be directed by an external magnetic field. Thus, the overall non-equilibrium self-assembly behavior
of this correlated system is tunable by a combination of light and magnetic fields. Another recent experimental study \cite{Lumay2015} reports
the onset of cooperative swimming of ferromagnetic dipole-coupled assemblies on a liquid-air interface due to a pulsating magnetic field. The locomotion is induced by periodic deformations
of the particle arrangement (rather than by capillary waves \cite{Snezkho2009}). Finally, we mention experiments on quasi-2D suspensions of dielectric patchy particles in {\em vertical} electric fields \cite{Sano2015}. These particles perform random, in-plane swimming motion while interacting via repulsive dipolar interactions. The experiments \cite{Sano2015} reveal mesoscopic turbulence similar to what has been seen in bacterial solutions.

From the theoretical side, Kaiser et al. \cite{Kaiser2015} have investigated cluster formation of active dipolar particles with permanent (magnetic) dipoles by means of numerical solution 
of overdamped, deterministic
(noise-free) equations of motion. Specifically, the active dipoles are represented by dipolar soft spheres with a propulsion force in the direction of the dipole moment. The self-propulsion is shown to create
complex cluster dynamics, involving a variety of states with different (non-equilibrium) internal structures and
different modes of motion of the cluster's center of mass.
\subsubsection*{Lane formation}
A special class of dipolar colloidal swimmers is realized by metallodiectric (gold-patched polystyrene) spheres in 
a 2D set-up with an {\em in-plane} AC (electric) field.
At certain frequencies below the critical one (for the high-frequency regime, see Sec.~\ref{networks}), experiments \cite{Gangwal2008_b} report
spontaneous motion of each particle in one of the two directions
orthogonal to the field, specifically, away from the particleÕs gold
patch. The underlying mechanism foots on the difference in strength of the (parallel) induced dipole moment in the metallic and dielectric part of the sphere. This asymmetry
yields an asymmetric flow
of solvent charges induced by the (in-plane) AC electric field (induced-charge electrophoresis). 

\begin{figure}
	\centering
	\includegraphics[width=\linewidth]{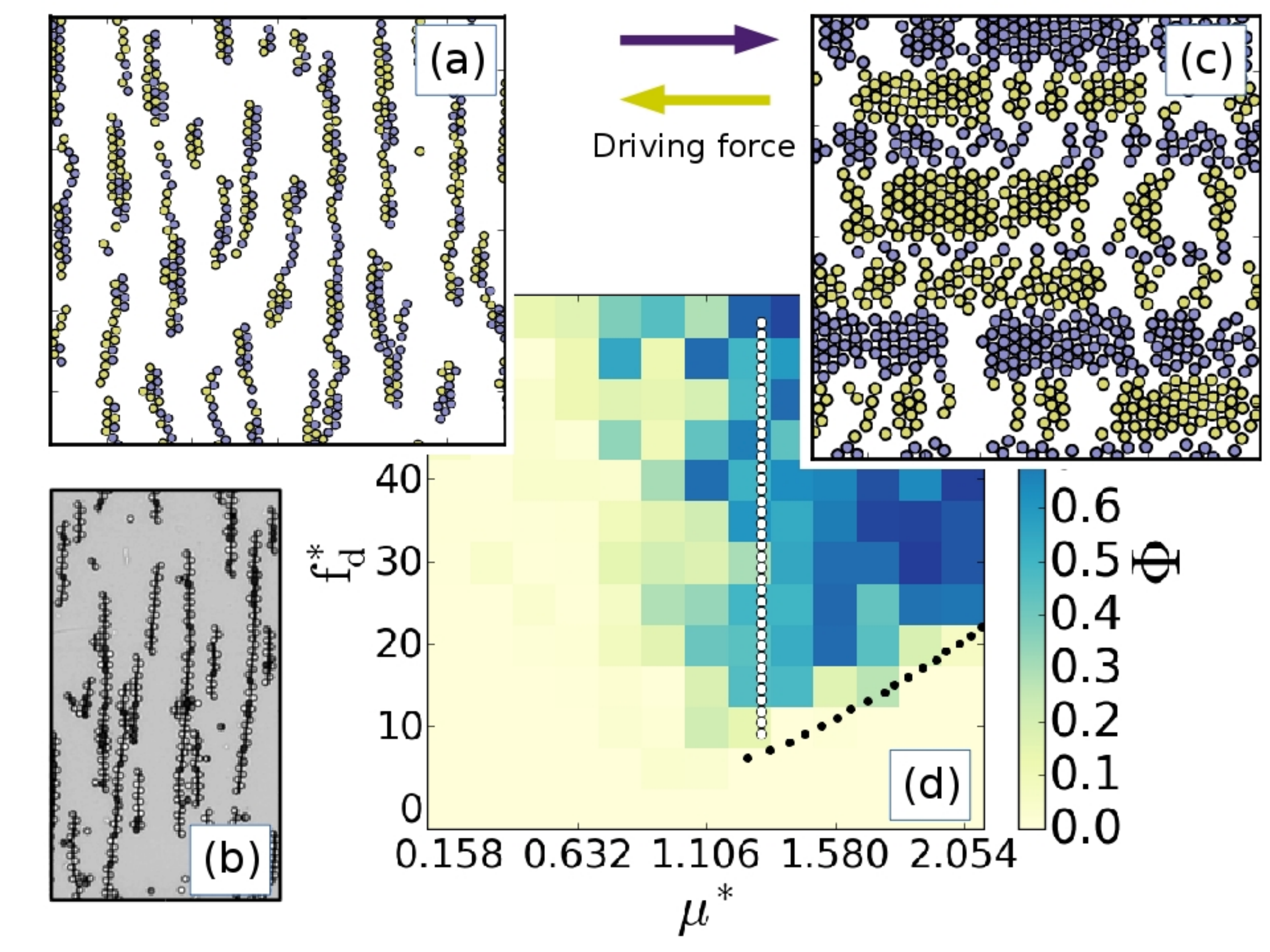}
	\caption{Lane formation in binary systems of dipolar microswimmers moving into opposite directions in response to a propulsion force $f_d$ (results from a BD simulation
	study \cite{Kogler2015_a}). (a) Assembly into staggered chains at zero force, consistent with experimental results (b) \cite{Gangwal2008}. (c) Snapshot of the laned state.
	(d) Laning state diagram at fixed particle density, showing the occurence of lanes (measured by the order parameter $\Phi$)  at sufficiently large driving forces and certain
	ranges of the dipolar coupling strength $\mu^{*}$. Vertical line: coupling strengths related to a spinodal instability (condensation transition) in the equilibrium system ($f_d=0$).
	Blue-dotted line: Estimate for the break-down of laning based on a balance between the attractive force stemming from multipolar interactions and 
	the driving force \cite{Kogler2015_a}).}
	\label{laning}
\end{figure}

The collective behaviour of these driven particles was investigated theoretically in \cite{Kogler2015_a} based on the asymmtric particle model sketched in Fig.~\ref{models}(e). 
Specifically, the study reports results from BD simulations of a binary 
mixture of particles with their smaller dipole moment either in the right or in the left part. The self-propulsion is modelled by a constant force whose direction
depends on the particle type, see Fig.~\ref{laning}.
The resulting equations of motion are given by $\gamma \dot {\boldsymbol r}_{i} = \sum^N_{j=1} \boldsymbol \nabla U(ij) + \boldsymbol f^s_{d,i} + \boldsymbol \zeta_i$
where $U(ij)$ is the full pair interaction, $\gamma$ is the friction constant, $\boldsymbol \zeta_i$ represent white noise, and $\boldsymbol f^s_{d,i}$ is the constant driving force.
At zero drive ($f_d=0$), the particles
assemble into staggered chains, which have been previously observed experimentally \cite{Gangwal2008}. Switching on the driving force, the system displays a transition
towards a laned state (Fig.~\ref{laning}(c)).
Indeed, lane formation is a protoype of a
non-equilibrium self-organization process, where an originally
homogenous mixture of particles (such as charged colloids) moving in opposite directions segregates
into macroscopic lanes composed of different species. As shown in \cite{Lichtner2015}, anisotropic dipolar interactions induce several novel features
within the (otherwise well-studied) laning transition. In particular, laning occurs
only in a window of interaction strengths, Fig.~\ref{laning}(d), which is closely related to phase separation process (as confirmed by density functional arguments \cite{Lichtner2015}). This is yet a further example demonstrating the intimate relationship between complex out-of-equilibrium collective behavior and equilibrium properties.
\section{Conclusions}
In the present article we have discussed recent advancements in understanding the collective dynamics of dipolar and multipolar colloids in and out of equilibrium. Theoretical progress in this emerging, very active field of research involves a broad variety of concepts from equilibrium statistical physics (targeting ground state structures, aggregation and phase transitions), from nonlinear dynamics of synchronizing
networks, and from the physics of pattern formation in driven or autonomous, dissipative systems, to name just a few.
At the same time, the field attracts substantial attention from material science in the context of developing
novel materials with "programmable" response to mechanical stress, shear, magnetic and thermal fields \cite{Bharti2015}, as well as  from microfluidics \cite{Tierno_review}. 
It is the interplay between these scientific communities which has led to the recent burst of attention in this area.

One obvious trend for future 
(theoretical and experimental) research is a shift towards the collective dynamics of {\em anisotropic} dipolar or multipolar particles, e.g., dumbbells, ellipsoids, or disks. Compared to spheres, shape anisotropy
can induce an enhanced sensitivity to external fields (regarding particularly rotational motion), but also additional degrees of freedom ("phase variables") in oscillating fields, enabling new types of synchronization and associated structure formation. Moreover, in the area of active systems, already 
non-polar shape-anisotropic particles display complex swimming modes not seen for spheres \cite{Hagen2013}. It would be very interesting to explore the swimming behavior of shape-anisotropic, dipolar swimmers which moreover offer additional ways of external guidance. Another promising direction concerns the {\em control}
of dynamical structures in such systems: Here one may envision the ãselectionÒ a specific dynamical structure (or mode of motion) from a pool of competing "candidates" by adapting the external field
based on information {\em from} the systems. Such feedback control \cite{Bechhoefer2005} strategies (which are well-established in the field of nonlinear science \cite{Schoell}) are already used for guiding (phoretic) swimmers \cite{Braun2015} and to manipulate dynamical structures \cite{Vezirov2015} and dynamical assembly of simple colloids \cite{Bevan2012}. Applications to the collective dynamics of dipolar and multipolar colloids thus seem very promising.

\section{Acknowledgement}
I would like to acknowledge collaborations and stimulating discussions with Sebastian J\"ager, Heiko Schmidle, Florian Kogler, Arzu B.~Yener, Ken Lichtner, Carol K. Hall and Orlin D. Velev. 
This work was supported by the Deutsche Forschungsgemeinschaft (DFG) through the International Research Training Group "Self-assembled soft matter nanostructures at interfaces"
IRTG~1524, the Research Training Group "Non-equilibrium collective phenomena in condensed matter and biological systems" RTG~1558, and the Collaborative Research Center
"Control of self-organizing nonlinear systems" SFB~910.

\end{document}